# Ionoacoustic detection of swift heavy ions

**S. Lehrack[a], W. Assmann[a,*], M. Bender[b], D. Severin[b], C. Trautmann[b,c], J. Schreiber[a], K. Parodi[a]**

a Department of Medical Physics, Ludwig-Maximilians-Universität München, Am Coulombwall 1, 85748 Garching, Germany

b GSI Helmholtzzentrum, Planckstraße 1, 64291 Darmstadt, Germany

c Technische Universität Darmstadt, Alarich-Weiss-Straße 2, 64287 Darmstadt, Germany

## Abstract

The maximum energy loss (Bragg peak) located near the end of range is a characteristic feature of ion stopping in matter, which generates an acoustic pulse, if ions are deposited into a medium in adequately short bunches. This so-called *ionoacoustic effect* has been studied for decades, mainly for astrophysical applications, and it has recently found renewed interest in proton therapy for precise range measurements in tissue. After detailed preparatory studies with 20 MeV protons at the MLL tandem accelerator, ionoacoustic range measurements were performed in water at the upgraded SIS18 synchrotron of GSI with $^{238}$U and $^{124}$Xe ion beams of energy about 300 MeV/u, and $^{12}$C ions of energy about 200 MeV/u using fast beam extraction to get 1 microsecond pulse lengths. Acoustic signals were recorded in axial geometry by standard piezo-based transducers at a 500 kHz mean frequency and evaluated in both the time and frequency domains. The resulting ranges for the different ions and energies were found to agree with Geant4 simulations as well as previous measurements to better than 1%. Given the high accuracy provided by ionoacoustic range measurements in water and their relative simplicity, we propose this new method for stopping power measurements for heavy ions at GeV energies and above. Our experimental results clearly demonstrate the potential of an ionoacoustic particle monitor especially for very intense heavy ion beams foreseen at future accelerator facilities.

\* **Corresponding author:** Walter Assmann, Department of Medical Physics, Ludwig-Maximilians-Universität München, Am Coulombwall 1, 85748 Garching, Germany. walter.assmann@lmu.de, phone: +49 89 28914283, office: +49 89 28914078

# 1. Introduction

First ideas to use thermoacoustic phenomena for particle detection date back to the experimental studies of Sulak et al. [1] and Askariyan et al. [2]. The technique has seriously been considered for underwater ultra-high neutrino detection for which appropriate detector arrays have been developed [3, 4]. There have also been attempts to use the acoustic signal induced by the characteristic dose deposition of an ion pulse in context of radiation therapy [5, 6]. Recently, this method has been reconsidered in advanced proton therapy, for which the so-called ionoacoustic signal promises a simple, but very accurate means to measure the Bragg peak position during patient irradiation (at least in favorable anatomical locations) [7]. Submillimeter range accuracy has been demonstrated in water [8, 9], and as an additional advantage, the ionoacoustic signal could be correlated with ultrasound imaging of the tumor morphology [10, 11]. Besides this medical application, an ionoacoustic particle detector also has great potential for monitoring intense proton or heavier ion bunches, as has been proposed in the early papers [1, 2]. For example, ionoacoustics offers a distinct detection technique for laser accelerated ions, which are produced in unique ultrashort bunches of high particle number accompanied by an interfering electromagnetic pulse (EMP). Acoustic detectors can take advantage of their huge dynamic range and moreover, the acoustic signal is separated from the EMP due to the longer transit time of the sound wave. This has recently been demonstrated for energetic protons accelerated by state-of-the-art PW class lasers, where the typically broad energy distribution of a single polyenergetic proton bunch was reconstructed using the ultrasound signal from a single piezo-composite (PZT) transducer [12]. Moreover, acoustic signals from GeV heavy ions have been studied at accelerators in various experimental configurations. At the RIKEN cyclotron (Japan), a PZT detector was used to investigate the creation mechanism and characteristics of acoustic waves generated by 95 MeV/u Ar ions in solid materials (Al, Cu, $BaF_2$) [13]. Also, a series of experiments using 400 MeV/u Xe ions delivered by the HIMAC synchrotron (Japan) explored the properties of various setups for acoustic detection of particles. Ions were stopped in different liquids as well as in the PZT-detector itself and the potential for using this ultrasound technique for heavy ion detection was stated [14, 15].

We report ionoacoustic measurements with GeV-ions from the upgraded SIS18 synchrotron at GSI (Darmstadt, Germany) where a water beam dump was exposed to short and intense heavy ion bunches (C, Xe, U) with energies from about 200 to 300 MeV/u. In contrast to most previous experiments, we used a single standard PZT-transducer in the axial configuration (i.e. on beam axis) to optimize the range and, hence, the energy resolution [16, 17]. Following a brief introduction concerning characteristic parameters of ionoacoustic signal generation with focus on heavy ions, extensive pre-studies with 20 MeV protons at the Munich tandem accelerator (MLL, Garching, Germany) are described along with methods for extracting the ion range from the measured acoustic signal pattern in the time and frequency domains. Determined heavy ion range values are compared to Monte Carlo simulations and, for C ions, also to existing data acquired with a more conventional technique, demonstrating the high accuracy of this ionoacoustic approach.

## 2. Ionoacoustic signal generation and simulation

Today, thermoacoustic methods are mainly used in opto- or photoacoustic applications, where local heating is induced by selective absorption of short-pulsed laser light [18]. The physical principles governing acoustic wave generation are similar for ionoacoustics differing only by the heating process and time profile [19]. The slowing down of energetic ions (of energy above several MeV) in matter is dominated by electronic excitation and ionization processes in the target material (electronic stopping). The dependence of stopping power $dE/dx$ on the ion velocity $v_{proj}$ and the velocity-dependent mean charge state of the ion $Z_{eff}$ is described by the well-known Bethe-Bloch formula with $dE/dx \propto Z_{eff}^2 / v_{proj}^2$, leading to an energy loss (depth dose) profile with a characteristic maximum (Bragg peak, BP) near the end of ion range. The microscopic processes occurring sequentially along an ion trajectory are complex: The primary ionization processes initiate an electron cascade, which radially distributes the deposited energy on an ultra-short time scale ($10^{-15}$-$10^{-13}$ s) around the ion trajectory. Within the following $10^{-13}$-$10^{-11}$ s thermalized electrons transfer energy to the atomic sub-system by electron-phonon coupling. In many solids, melting and quenching occurs on a sub-nanosecond timescale resulting in the formation of a few nanometer-wide ion track. In the case of water, local temperatures can exceed the boiling point near the BP. The whole process has been quantitatively described by an inelastic thermal spike model [20] and, except for the very first stage, even Coulomb explosion has been discussed [21].

Considering such ion heating on the macroscopic scale, we define for ionoacoustics a characteristic Bragg peak volume (BPV) by its axial dimension being the distance between the two turning points of the Bragg curve before and after the BP maximum, and the lateral dimension as the full width at half maximum (FWHM) of the beam spot size. Assuming a typical BPV of 0.25 cm$^3$ and a short ion bunch of order 10$^6$ ions (including heavy ions) delivered to a water volume would increase the local temperature by no more than 1 mK. Nevertheless, even such a low temperature increase can generate a pressure pulse $\Delta p$, in accordance with $\Delta V/V = -\kappa \Delta p + \beta \Delta T$ (with $\kappa$ being the isothermal compressibility coefficient and $\beta$ being the volume expansion coefficient). The maximum pressure will be reached, if the following two conditions are fulfilled: (i) the energy deposition is adiabatic (thermally confined), i.e. faster than the rate of thermal energy diffusion from the BPV, which is of order milliseconds for water, and (ii) the energy deposition is isochoric (stress confined), therefore it must be more rapid than the corresponding expansion rate of the BPV. Determined by the sound velocity in water ($c \approx 1.5$ mm/µs) and the size of the BPV, this typically limits the maximum beam bunch duration at high ion energies to a few microseconds. In thermal and stress confinement, $\Delta V = 0$ and the pressure pulse $\Delta p$ due to a temperature rise $\Delta T$ can therefore be estimated according to $\Delta p = \beta \Delta T / \kappa$, which amounts to about 400 Pa/mK (4 mbar/mK) in water. The general equation for generation and propagation of the thermoacoustic wave in space and time (see e.g. Ref. [22]), from which the pressure signal at a certain detector position **r** can be calculated, reduces in this case to:

$$p(\boldsymbol{r}, t) = \frac{\beta}{4\pi C_p} \partial_t \int d\boldsymbol{r}' \frac{H_s(\boldsymbol{r}') H_t(t - \frac{|\boldsymbol{r} - \boldsymbol{r}'|}{c})}{|\boldsymbol{r} - \boldsymbol{r}'|} \tag{1}$$

using a heating function $H(\boldsymbol{r}', t)$ defined as the ion induced thermal energy input at $\boldsymbol{r}'$ and at time $t$. This can be separated here into independent spatial and temporal contributions: $H_s(\boldsymbol{r}')$ and $H_t(t)$ (for details see Ref. [19]). The spatial part of the heating function $H_s(\boldsymbol{r}')$, given by the energy loss of the ion in the stopping medium, can be evaluated using *Geant4* [23]. For this calculation, we used version 10.01.p02 with the QGSP_BIC_EMZ (EM option 4) physics list for the main electromagnetic and nuclear processes. The value of the ionization potential of water was set to 78 eV according to both the recommendation of Sigmund et al. [24] and best fits to our recent experimental results [25, 8]. The spherical integral given in Eq. (1) was solved by an analytical approach using for $H_s$ a dose distribution according to Ref. [26]. Alternatively, as a second approach, a simulation of the thermoacoustic wave generation and propagation was performed via the pseudospectral partial differential equation solver in the MATLAB toolbox *k-Wave* [27]. The required temporal pulse profile $H_t$ is then convolved in a second step influencing also the frequency spectrum of the signal [19]. The mean (carrier) frequency, $f_{mean}$, of the ultrasound signal is determined by the BP width, $l$ ($f_{mean} \approx c/l$), the shape of the Bragg curve and the ion pulse duration, as well as the spectral shape and in particular the higher frequency components of the signal. While the first analytical approach can be considerably faster for single point detectors, the second offers the possibility to include heterogeneous structures and specific detector shapes. Ideally, for a complete evaluation of the measured ultrasound signal the transducer specific transfer function (TIR) has to be considered, which must include geometrical effects as well as the frequency depending sensitivity [28].

## 3. Ionoacoustic setup and proton test experiments

In all the presented measurements we used our standard setup (Fig. 1) [8]: Ions left the beam line vacuum through an exit window (for protons 11.4 μm Ti foil) and entered via 80 mm long air-filled entrance channel a water container (33 × 18 × 19 cm$^3$) through a polyimide entrance window (for protons of 50 μm thickness). During the experiments the temperature of the deionized water was continuously measured with an accuracy of ±1 °C to later correct the temperature-dependent speed of sound. Acoustic signals were recorded by ultrasound immersion PZT transducers (Videoscan, Olympus) mounted on the incident ion beam axis at an appropriate distance from the calculated BP position (see Fig. 1). As shown in Ref. [19], axial positioning of the transducer enables the most accurate range determination. Size, shape (i.e. focused or unfocused) and detector frequency range were chosen according to the ionoacoustic requirements. Signals were amplified with a low-noise broadband amplifier (Miteq AU1213) and recorded with a digitizing oscilloscope (Rhode & Schwarz RTM2034) typically at a 500 MS/s sampling rate.

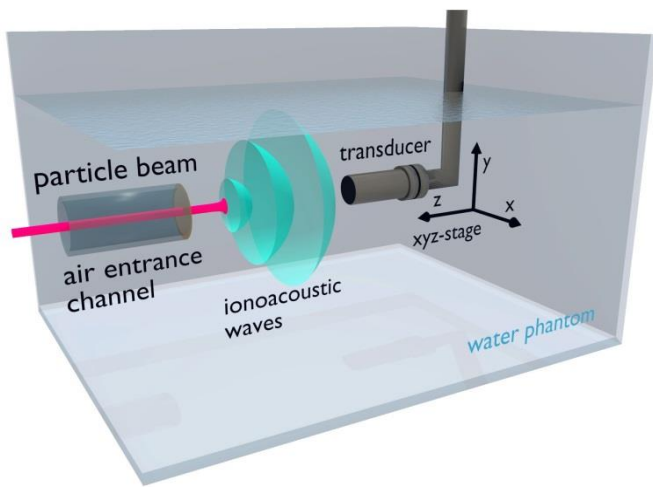

**Fig.1**: Standard setup for ionoacoustic experiments: Ions enter a water filled acrylic container through an air filled channel and a thin Kapton entrance window, are slowed down and finally stopped at the Bragg peak. The induced ultrasound waves propagate towards a PZT-transducer mounted on a remote controlled xyz-stage (Fig. 1 from Ref. [8]).

In order to test this relatively novel particle detection technique and to study the achievable range and energy resolutions, a series of proof-of-principal experiments were performed using 20 MeV protons delivered by the Munich electrostatic tandem accelerator at MLL (Garching, Germany) with an energy uncertainty $\Delta E/E$ near $10^{-4}$. The position of the BP maximum in water was calculated by Geant4 to be 4.03 mm, taking into account the energy loss of the protons in the vacuum exit foil, the air gap and the polyimide entrance foil. The BP width defining in axial configuration the relevant length for stress confinement condition was estimated to be 0.3 mm. Making use of the MLL chopper system, which can deliver a wide range of pulse lengths, a rectangular ion bunch with a 3 ns rise/fall time and duration less than 200 ns was accordingly selected. The chopper electronic delivered also a precise trigger for the digitizing oscilloscope allowing signal averaging, which was necessary due to the low signal-to-noise-ratio (SNR). A typical 125-fold averaged ultrasound signal of a 110 ns bunch of about $10^6$ protons with a lateral BP area of about 1 $mm^2$ is shown in Fig. 2. In this case, a PZT transducer (of 3.5 MHz central frequency and 60 dB amplification) was positioned on beam axis at the detector's focal distance of 25 mm behind the BP position. It should be noted, that the measured signal corresponds to a pressure near 5 Pa (0.05 mbar) at the transducer position.

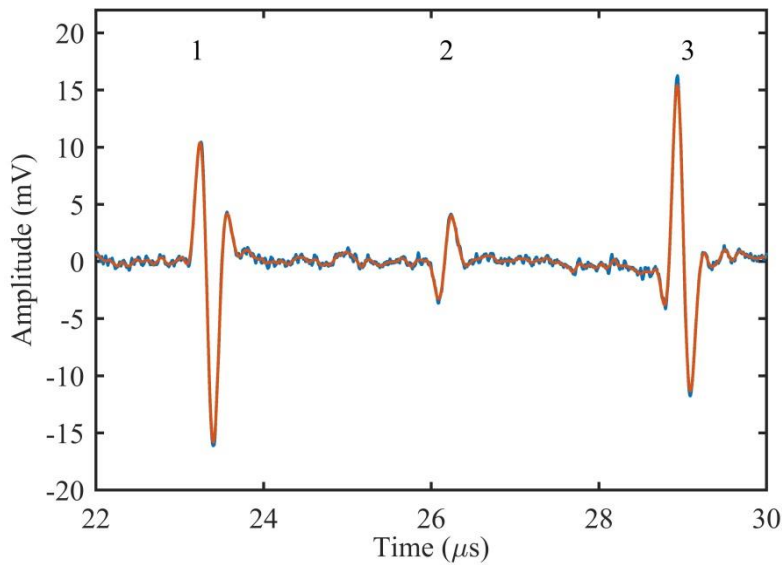

**Fig. 2**: Ionoacoustic signal of a 20-MeV proton beam with $1.3\times10^6$ protons per pulse and 110 ns pulse length, amplified with 60 dB and 125-fold averaged, for further explanation of numbers see text (raw data in blue, smoothed data in red).

## 4. Signal evaluation in the time and frequency domains

The numbered peaks in Fig. 2 can be assigned to the first arriving ultrasound signal (1) from the BPV itself, to a later signal (2) produced at the entrance window (polyimide foil) and to a last signal (3) due the reflection of the BPV signal from the entrance foil, where it encounters a 180 degree phase shift. In stress confinement, the typical bipolar signal shapes are the result of a superposition of contributions from the spatial and temporal heating functions. The different signal components are better distinguished in Fig. 3, where the proton bunch duration is varied from 55 to 1030 ns, i.e. beyond stress confinement, showing a clear separation of the spatial ($H_s$) and temporal ($H_t$) contributions. Fundamentally, the pressure signals are not due to heating or cooling, but generated by the temporal dynamics of the absorbed energy or dose, i.e. more notably at the ion bunch rise and fall times, in accordance with Eq. (1). The spatial heating (dose depth) profile is assumed to be instantaneously established in the stopping medium, during the early part of the ion bunch and within the approximate sub-nanosecond time interval while particles traverse the medium and are stopped. Its contribution to the ionoacoustic signals (1) to (3) are fully separated at 1030 ns bunch length, ordered according to their arrival time at the detector (see Fig. 3).

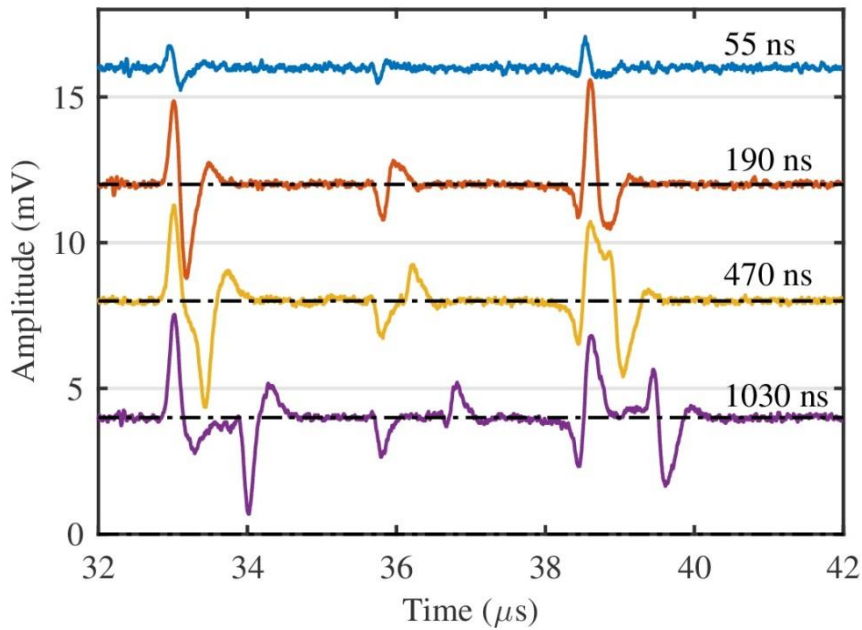

**Fig. 3**: Ionoacoustic signals of a 20 MeV proton beam with different pulse length.

The first arriving component of signal (1) exhibits a positive pressure (compression) peak from the BPV front side (i.e. downstream ) followed by a broader negative (rarefaction) peak of lower amplitude generated by the BPV backside (i.e. upstream). The signal shape mirrors the first order spatial derivative of the underlying heating profile with its steeper slope at the front side (see Fig. 6). The second part of signal (1) arriving 1030 ns later (near 34 µs in Fig. 3) has an identical shape with opposite sign due to the temporal extinction of energy deposition at the end of the bunch. This latter part of signal (1) therefore begins with the ionoacoustic signal from frontside of the BPV followed by that from the backside. The first part of signal amplitude (2) is as expected negative due to the abrupt impedance change from air to water, marginally influenced by the polyimide entrance foil. Signal (3) is like a mirror image of signal (1) including a phase change due to reflection: the compression signal from the BPV backside arriving first, then the rarefaction signal from the front side arriving later. The fact that the measured raw signal corresponds so closely to the expectation from Eq. (1) is due to the flat (and therefore in our case negligible) spectral response curve of the transducer setup.

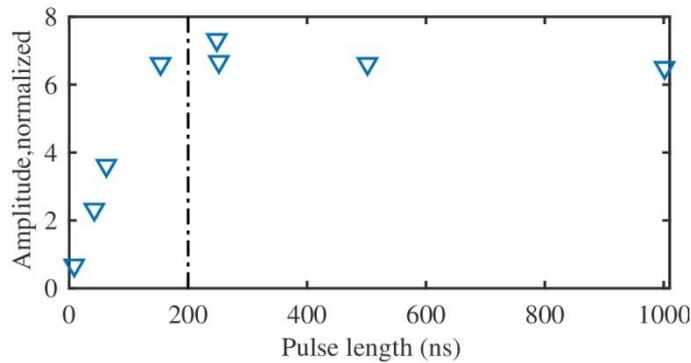

**Fig. 4**: Signal amplitude at Bragg peak maximum for varying proton pulse length. The vertical line represents the stress confinement time for the particular case of 0.3 mm Bragg peak width.

It is interesting to note that the stress confinement condition also influences the signal amplitude as is displayed in its dependence on the proton bunch length of Fig. 4, where the pulse length is increased from 8 ns to 1 µs at constant ion current before the chopper. Inside stress confinement, in agreement with Ref. [1], the signal amplitude raises with growing number of ions, but for bunch length beyond about 200 ns (i.e. outside stress confinement for 0.3 mm BP width) meaning that further heat input has no effect on the generated pressure amplitude.

The well-defined conditions in the proton experiments allowed also tests, both in time and frequency domains, of evaluation techniques for determining the position of the BP maximum (i.e. location of the maximum dose). For the sake of simplicty, we use in the following the term 'range' interchangeably with BP maximum position. However, it has to be emphasized that the term 'range' is more rigorously defined as the location at a specified dose fraction (typically 80 %) of the BP maximum on the downstream side. Under stress confinement, the time difference between signals (1) and (2), multiplied with the appropriate sound speed, gives directly the proton range in water, and similarly the time difference between signals (1) and (3) is twice this range. According to Eq. (1), the signals are induced by the temporal gradient of the heating function $H_t$, therefore, the zero-crossing of the signals corresponds to the BP maximum position within the BPV. The accuracy of this method depends on the proper determination of the zero-crossing point as well as on the temperature corrected speed of sound, which we determined from the measured water temperature and a fit to experimental data by Ref. [29]. Alternatively, the actual sound speed can also be measured by varying the transducer axial position in precise steps and, from the corresponding time shifts in the signal, the speed of sound can be directly calculated. Sound speeds determined in these test experiments with both methods agreed within 0.1 %. The calculation method using a fit function from Ref. [29] is somewhat easier to perform, but depends critically on the accuracy of the temperature measurement. A detailed study of other possible metrics that can be extracted from the proton BP position based on the pressure signal arrival time, even at pulse lengths outside stress confinement, can be found in Ref. [19].

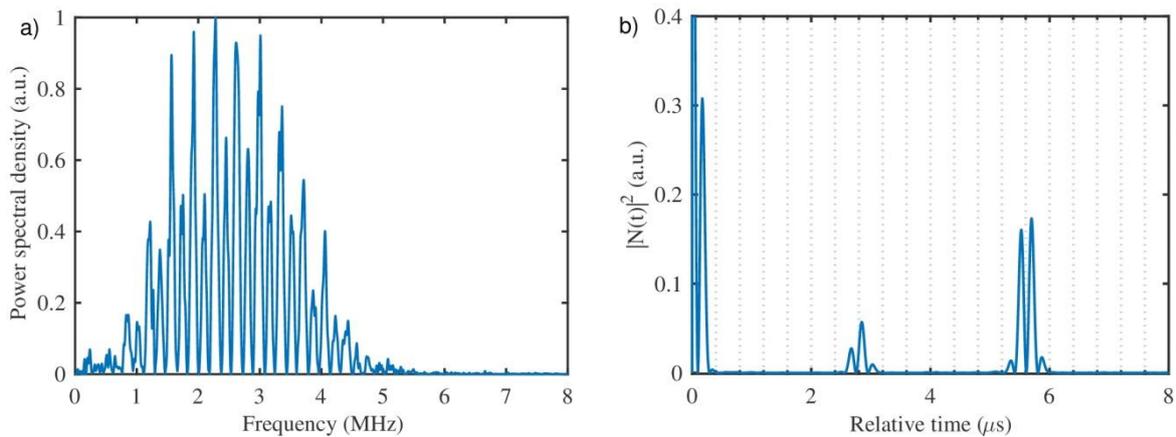

**Fig. 5**: Spectral power density $|F(S(t)|^2$ (a) and autocorrelation spectrum $|N(t)|^2$ (b) of the ionoacoustic signal shown in Fig. 2.

Selection of the right zero-crossing points in Fig. 2 is straightforward, however, phase shifts due to different TIR can significantly change the picture and can therefore complicate the zero-crossing determination, to the same extent as would superimposed complex temporal profiles. For low pressure pulses, low SNR can additionally hamper the range determination in the time domain, in spite of signal averaging. In the frequency domain, one can make use of the fixed correlation between signals (1), (2) and (3) determining the autocorrelation function *N(t)* (according to the Wiener-Khinchin theorem [30]) by inverse Fourier transformation $\boldsymbol{F}^{-1}$ of the power spectrum relevant to signal *S(t)*:

$$N(t) = \boldsymbol{F}^{-1} \{|\boldsymbol{F}[S(t)]|^2\} \qquad (2)$$

This method is less sensitive to noise and is demonstrated exemplarily for the spectrum shown in Fig. 2: The power spectrum is displayed in Fig. 5a and the corresponding absolute value of the autocorrelation spectrum in Fig. 5b. One can clearly observe two double-peaks (due to the bipolar signals) centered around the expected flight time of about 2.7 µs (window signal) and 5.4 µs (reflection signal), the distinct minima correspond to the zero-crossings in time domain. The window signal is generally weaker than the reflection signal, therefore we limit further analysis to the latter. Table 1 summarizes results from several of our proton experiments under slightly different experimental conditions, evaluated in the time domain (zero-crossing of reflection signal) as well as in the frequency domain (autocorrelation minimum). Range values are given in terms of differences between experimental and simulated results as no other range measurements are available for comparison in the literature at this energy. While the reflection results agree within ± 2% with Geant4, autocorrelation ranges exhibit a small constant positive offset and, thus, the zero-crossing determination of the time signal seems to be slightly more accurate than the autocorrelation method for range determination. If only range (or energy) differences are of importance, however, this offset is irrelevant and both methods have a comparable accuracy near 1%.

| Exp | Geant4 range (µm) | Δ $_{Auto}$ (µm) | σ $_{Auto}$ (µm) | Δ $_{Reflection}$ (µm) | σ $_{Refl}$ (µm) |
|---|---|---|---|---|---|
| 1 | 4028 | 60 | 13.0 | -79 | 17.2 |
| 2 | 4030 | 56 | 2.8 | -2 | 5.3 |
| 3 | 4031 | 118 | 21.5 | 56 | 6.3 |
| 4 | 4034 | 102 | 29.3 | 35 | 41.7 |
| 5 | 4034 | 72 | 92.6 | -13 | 8.7 |
| 6 | 4424 | 139 | 64.3 | 66 | 19.3 |

**Table 1:** Compilation of experimental data with 20-MeV protons: Difference Δ of experimental mean values from two evaluation methods (reflection peak or autocorrelation analysis) to Geant4 ranges, σ is the standard variation for each evaluation method. (see also text).

## 5. Experimental results

Heavy ion experiments were performed at the upgraded SIS18 synchrotron (GSI) with ion beams of $^{12}$C (180– 240 MeV/u), $^{124}$Xe (280 – 320 MeV/u) and $^{238}$U (250 – 300 MeV/u) with an energy resolution, ΔE/E of $10^{-3}$. Exemplary energy deposition (Bragg) curves from Geant4 simulations in water are presented in Fig. 6 for ion energies of 200 MeV/u ($^{12}$C ) and 300 MeV/u ($^{124}$Xe and $^{238}$U). At the utilized energies, ion ranges vary from 71 to 117 mm for $^{12}$C and from 10 to 24 mm for the heavier ions where the typical BP width can vary from 1.0 to 2.5 mm. To be in stress confinement, fast beam extraction was used delivering an ion macro-bunch of about 1 µs with a micro-structure consisting of 4 - 6 ion bunches with 100 ns bunch duration and, in addition, a precise trigger pulse for data recording. The number of ions in a macro-bunch was varied between $10^4$ and $10^6$ within a beam spot diameter of several millimeters. In some measurements with $^{238}$U, the number of ions per macro-bunch was below the beam line monitor threshold of $10^4$ particles and estimated, therefore, by linear extrapolation. The standard setup described above for the experiments with protons was only different at the beam entrance: the vacuum exit foil of the vacuum beam line was a 100 µm stainless steel foil, and the water entrance foil was replaced by a 300 µm polyimide foil. Between these two window foils, ions traversed an air gap of 60 to 65 cm. Due to the spot size, the BP width of order several millimeters and the relatively large distance between the entrance window and the BP, unfocused PZT transducers of 500 kHz central frequency were used with 60 dB signal amplification.

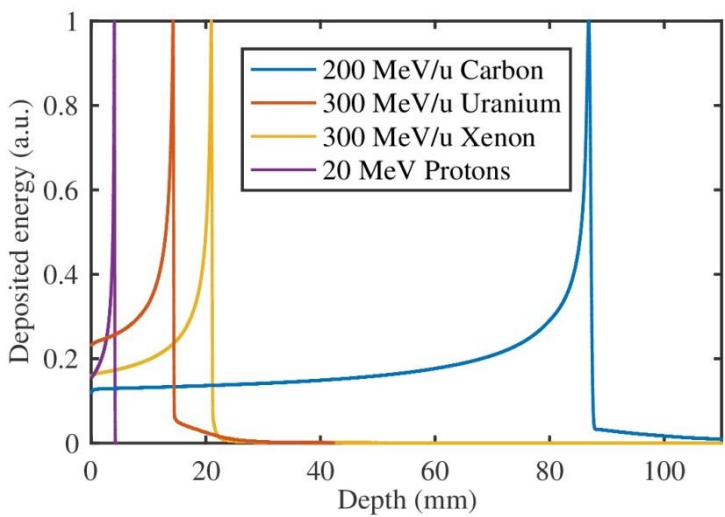

**Fig. 6**: Geant4 simulation of normalized energy deposition in water for protons (20 MeV), $^{238}$U (300 MeV/u), $^{124}$Xe (300 MeV/u) and $^{12}$C (200 MeV/u) (left to right).

## 5.1 Measurements with $^{238}$U and $^{124}$Xe ions

Examples of ionoacoustic spectra from 300 MeV/u $^{238}$U and $^{124}$Xe ions measured with a 500 kHz transducer (Fig. 7) display a clear acoustic signal pattern resembling that obtained with protons but also superimposed by the synchrotron micro-structure. Moreover, due to the higher energy loss compared to protons, the increased pressure signal amplitudes enabled single pulse measurements without averaging. Notably, the example spectrum of $^{238}$U in Fig. 7a exhibits an ionoacoustic signal from a single macro-bunch estimated to contain only 200 ions. The more pronounced micro-structure for $^{124}$Xe ions seen in Fig. 7b mirrors their smaller BP width. Due to the complex ionoacoustic signal (micro/macro-)structure the autocorrelation method lent itself in this case to range determination as demonstrated in Fig. 8 for $^{238}$U ions.

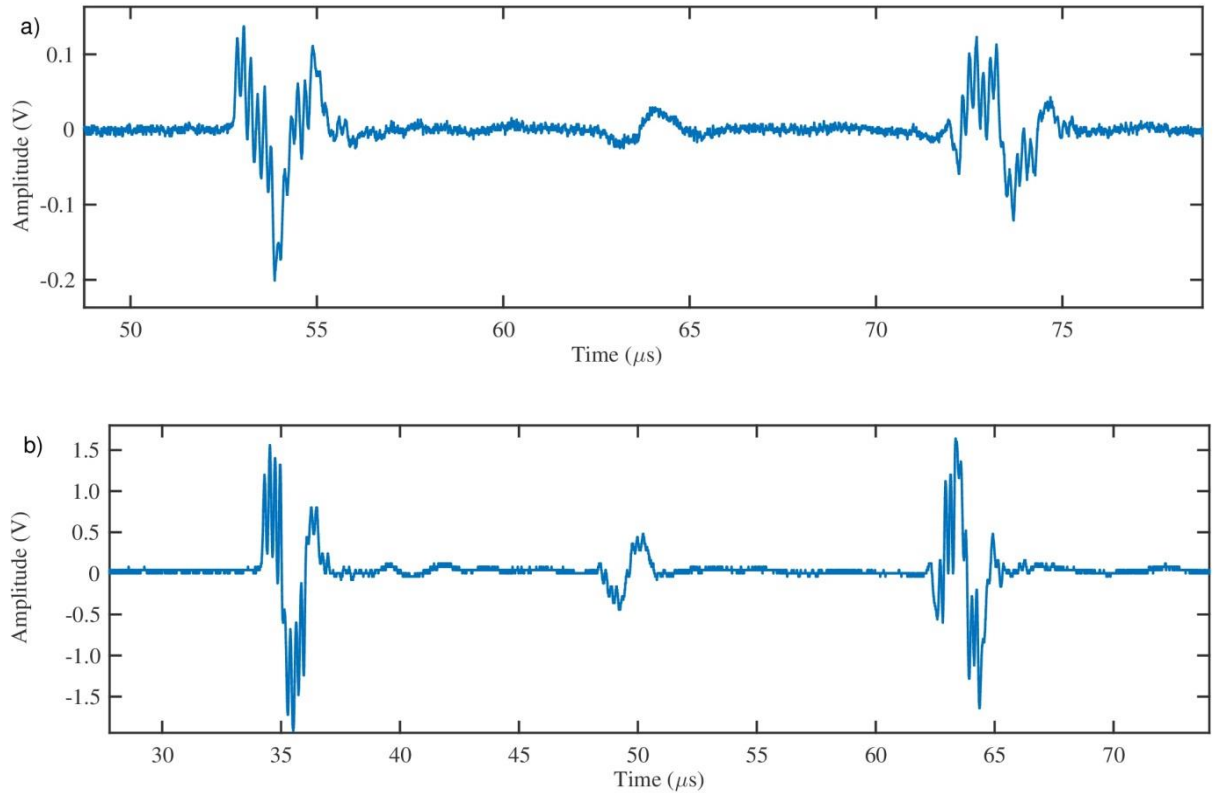

**Fig. 7**: Ionoacoustic signals of (a) 300-MeV/u $^{238}$U of about 200 ions per pulse, and (b) 300-MeV/u $^{124}$Xe with $10^6$ ions per pulse (see text).

A distinct feature can be taken from the power spectrum of Fig. 8a: It is dominated by the main signal frequency around 500 kHz (as expected from the BP structure) and is matched to the mean frequency of 500 kHz of the transducer and its frequency bandwidth of 80%. At higher harmonics of the transducer, frequency components of the signal are also apparent and most pronounced near 5.4 MHz, the SIS18 extraction frequency for 300 MeV/u $^{238}$U ions.

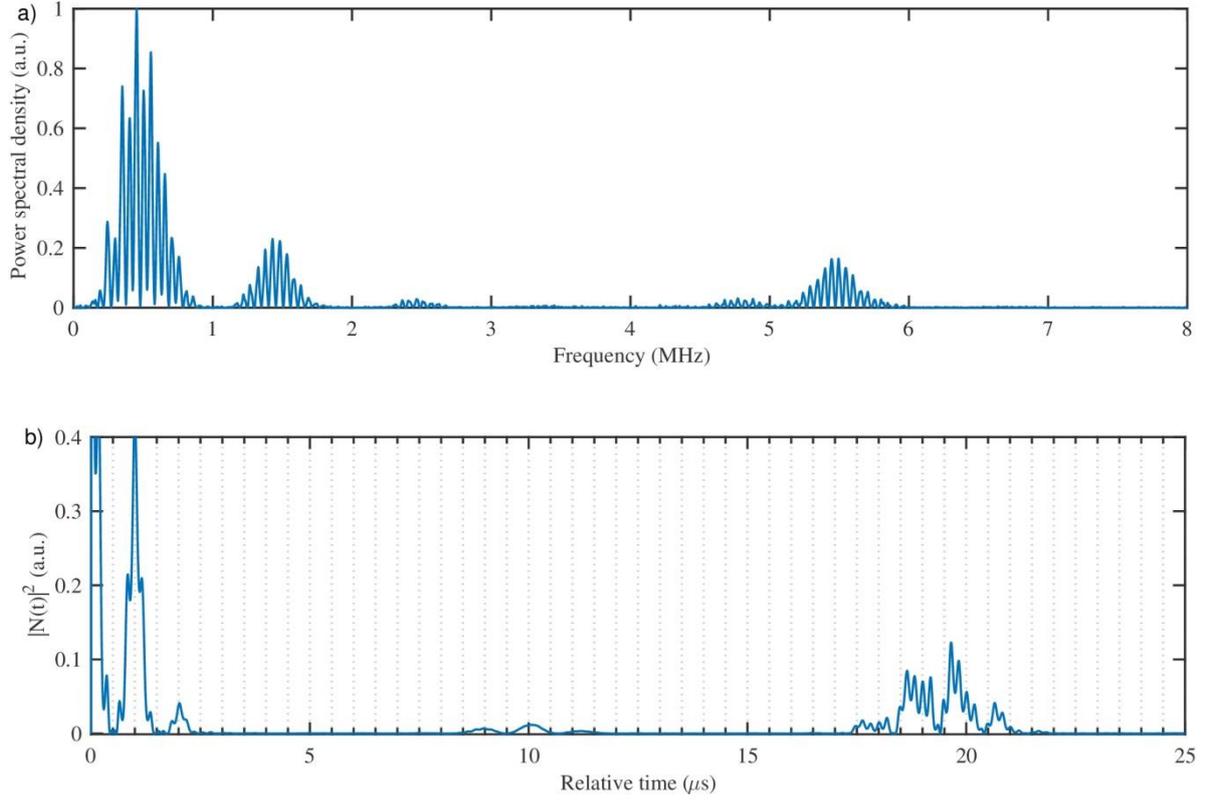

**Fig. 8**: Spectral power density $|F(S(t))|^2$ (a) and autocorrelation spectrum $|N(t)|^2$ (b) of the ionoacoustic signal from $^{238}$U ions shown in Fig. 7a.

Autocorrelation analysis shown in Fig. 8b yields a clear signal near 20 µs corresponding to twice the ion range. Additionally for comparison, we investigated a different time analysis for range determination using defined threshold values for the direct and reflected BP ultrasound signal. Table 2 presents a comparison of these experimental range values (which are mean values of 20 to 50 consecutive measurements) to simulated range values from Geant4 taking into account the specific ion energy loss prior to water entrance. As with Table 1, the difference between values from both evaluation methods and the corresponding Geant4 range values is listed. The precision of the two evaluation methods can be estimated from the standard deviation σ of subsequent measurements at the same energy. For the accuracy of the given range values, different contributions to the systematic errors were considered: The simulation error in Geant4 consists of the geometrical uncertainty of the experimental setup, the mean ionization potential and the temperature of water. Reasonable variations of these values result in a total simulated range error of 32 µm. The uncertainty of the experimental ranges is dominated by the temperature precision of ±1K, relevant to the speed of sound, which gives a range uncertainty of 42 µm for $^{124}$Xe ions and 29 µm for $^{238}$U ions. The agreement of both measured and simulation values is within this uncertainty range to better than 1%. Taking into account, that several measurements at the same energy were performed non-consecutively at different times, experimental $^{124}$Xe data extracted from autocorrelation analyses are in remarkable agreement with simulations. Additionally, the autocorrelation method can be facilitated, if the approximate minimum position is determined by the centroid of the correlation peaks as a first step. As mentioned above for protons, the speed of sound can be determined in two different ways, which was also tested with heavier ions. An axial scan is shown in Fig. 9 for $^{124}$Xe ions, which yields a sound velocity consistent within 0.5% with a fit based calculation according to Ref. [29]. In another experiment with $^{124}$Xe, the particle number was varied within the $10^6$ ions/pulse range looking for the corresponding dependency of the measured pressure amplitude (Fig. 10), and the smallest RMS error was achieved by a linear fit to the data.

| Ion | Energie (MeV/u) | Geant4 range (mm) | Δ Refl (μm) | σ Reflection (μm) | σ Reflection (keV/u) | Δ Auto (μm) | σ Autocorrelation (μm) | σ Autocorrelation (keV/u) |
|---|---|---|---|---|---|---|---|---|
| $^{12}$C | 180 | 70.75 | -80 | 7.8 | 11.6 | -460 | 7.0 | 10.4 |
| | 200 | 85.58 | 300 | 59.4 | 80.6 | -40 | 41.1 | 55.8 |
| | 220 | 100.84 | 140 | 55.8 | 71.3 | -180 | 33.2 | 42.4 |
| | 240 | 117.07 | 310 | 7.6 | 9.0 | -70 | 6.5 | 7.7 |
| $^{124}$Xe | 280 | 18.60 | 1 | 4.6 | 40.5 | -46 | 1.5 | 13.2 |
| | 290 | 19.78 | 31 | 4.9 | 43.4 | -21 | 1.4 | 12.4 |
| | 300 | 20.96 | 38 | 8.4 | 71.2 | -4 | 1.7 | 14.4 |
| | 310 | 22.18 | 62 | 4.1 | 34.9 | 25 | 1.7 | 14.4 |
| | 320 | 23.41 | 82 | 4.3 | 36.6 | 47 | 1.7 | 14.4 |
| $^{238}$U | 250 | 10.41 | -65 | 8.3 | 115.4 | 87 | 26.4 | 367 |
| | 280 | 12.69 | -35 | 14.7 | 187.5 | 101 | 25.4 | 324 |
| | 290 | 13.48 | -17 | 6.4 | 79.7 | 93 | 21.6 | 269 |
| | 300 | 14.29 | -10 | 13.3 | 156.1 | 49 | 11.5 | 135 |

**Table 2:** Compilation of experiments with $^{12}$C, $^{124}$Xe and $^{238}$U: Difference Δ of measured mean values from two evaluation methods (reflection peak or autocorrelation analysis) to Geant4 ranges, σ is the standard variation for each evaluation method in μm and converted to keV/u (see also text).

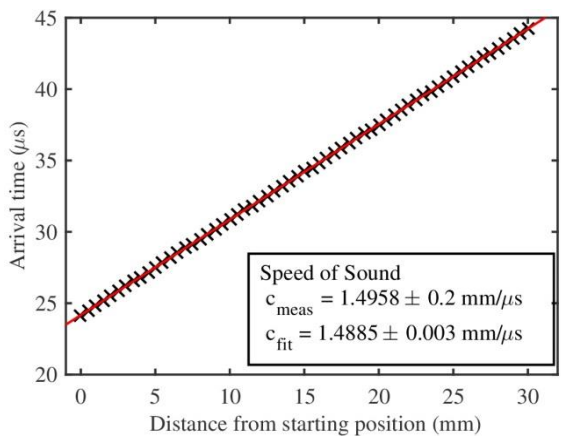

**Fig. 9**: Ionoacoustic signal of 300-MeV/u $^{124}$Xe, with varying Bragg peak-detector axial distance (axial scan) for speed of sound determination, compared to a fit on data from Ref. [29] (red line).

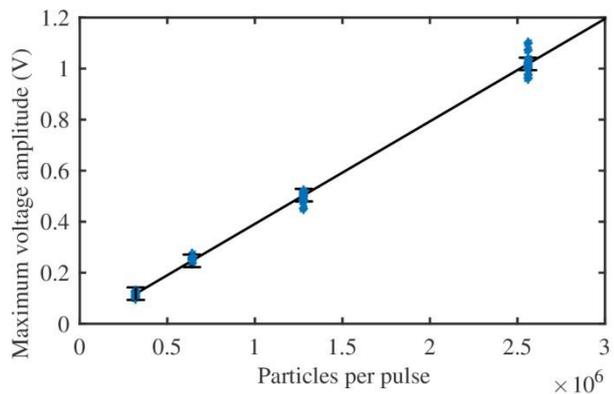

**Fig. 10**: Dependency of pressure signal amplitude of $^{124}$Xe ions (300 MeV/u) on particles per pulse compared to a linear fit.

## 5.2 Measurements with $^{12}$C ions

Ionoacoustic experiments with $^{12}$C ions differ in several aspects from the experiments with heavier ions. Due to their lower mass the range of $^{12}$C ions at different energies is on average 5 times longer. Thus longitudinal energy loss straggling is more pronounced leading to increased BP width, hence, signals with correspondingly lower mean frequency, and blurring of the beam micro-structure to less complex signal shapes as displayed in Fig. 11. Therefore, data evaluation could be performed with similar precision in the time and frequency domains. Geant4 ranges are compared in Table 2 with measured deviations of both evaluation methods (as described in 5.1). The agreement with Geant4 ranges for both methods is considerably better than 1%, including data from consecutive and non-consecutive repetition measurement.

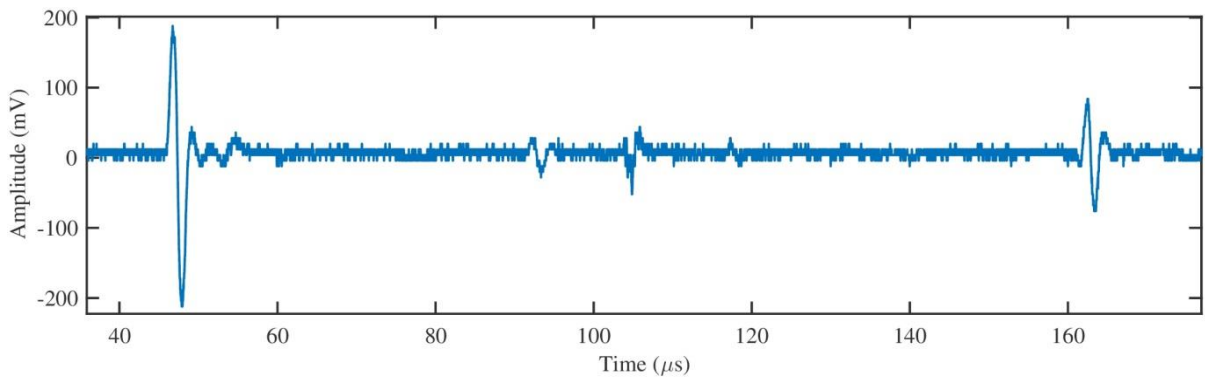

**Fig. 11:** Ionoacoustic signal of 200-MeV/u $^{12}$C with $1.0\times10^8$ ions per pulse and 1 μs pulse length.

It should be noted that the experimental error due to the temperature uncertainty is increased to about 175 μm on account of the larger range. Here, our results can be compared to earlier range measurements, which had been performed at GSI with $^{12}$C ions at similar energies in the context of the development of tumor therapy with light ions [31]. In this case, a completely different setup was used consisting of a water column (WS) of precisely variable length in combination with parallel-plate ionization chambers (IC) at both ends in order to normalize the exit IC data with the entrance IC. A power fit was made on the WS data, allowing interpolation of these WS range values for energies utilized in our ionoacoustic experiments. Furthermore, the WS range results had been corrected for the upstream energy loss from beam exit to water entrance, i.e. the measured range values were extrapolated to the indicated incidence energies, therefore Geant4 and ionoacoustic values were corrected as well. In Table 3 the Geant4 ranges are listed for the indicated incident energies, the differences between the herein corrected ionoacoustic range (AC* from autocorrelation analysis) and the WS range, as well as the AC* and Geant4 range difference. The deviations are less than 1%, therefore good agreement can be claimed between the two different range measurement techniques, taking into account the experimental errors and necessary extrapolations and interpolations.

| Energy (MeV/u) | Geant4 range (mm) | Geant4 – AC* (µm) | WS – AC* (µm) |
|---|---|---|---|
| **180** | 72.45 | -530 | 345 |
| **200** | 86.81 | 50 | 600 |
| **220** | 102.15 | -180 | 20 |
| **240** | 118.39 | 70 | -180 |

**Table 3:** Range values of $^{12}$C ions in water: Comparison of water-column (WS) and ionoacoustic (AC*, autocorrelation analysis) experiments, both corrected to the given incident energy, together with calculated Geant4 values (see also text).

## 6. Discussion and Outlook

The information delivered by this ionoacoustic technique is the ion range in water, but usually the ion energy is of interest and measured with particle detectors. To deduce energy values from this method, an range-energy calibration should be performed for a specific detector setup. Using Table 2 data such a calibration is presented for $^{124}$Xe in Fig. 12, where measured values are shown together with a power fit of calculated ranges from Geant4. It can be noted, that range values from the time-based evaluation agree within 0.5 % with simulated ranges. This excellent agreement in turn confirms the latest ICRU recommendation of 78 eV for the value of the ionization potential of water [32]. The accuracy of absolute ion energies depends on the accuracy of the energy values specified by the accelerator, which was better than 0.1 % for SIS18. If only small ion range or energy changes are to be measured, absolute energy values are of less importance and the energy resolution of the ionoacoustic method has to be considered. This resolution can be estimated from the standard deviation σ derived from a normal distribution fit of consecutively measured range values at a certain energy. These σ values from time distance as well as autocorrelation evaluation are given in Table 2 in μm together with their conversion into an energy uncertainty given in keV/u using a fit to simulated incidence energy changes according to the range variations, which results in a remarkable energy resolution of $dE/E \leq 10^{-3}$. Each of the two range evaluation techniques have their specific advantages: The autocorrelation method uses an "objective" criterion and can easily be incorporated into a data evaluation procedure. In contrast, the time evaluation often needs to be checked by hand, but can, however, deliver more accurate range values. The choice of technique depends on the particular pressure signal shape, and as seen for $^{12}$C in Table 2, similar results can be achieved for clear signal patterns.

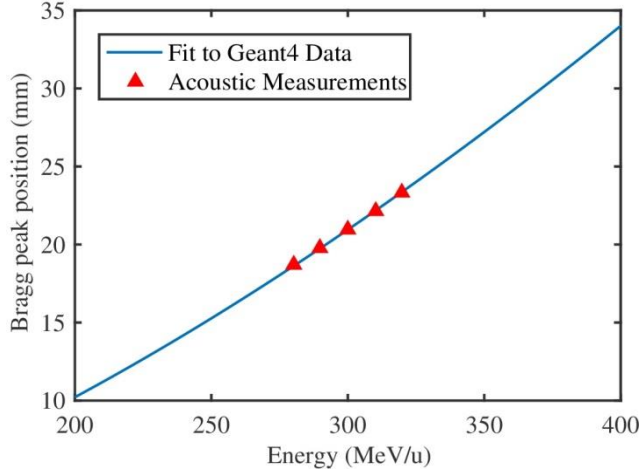

**Fig. 12:** Bragg peak position of $^{124}$Xe ions in water for different energies (red triangles) and a power fit on Geant4 values (blue line).

Another noteworthy aspect of the demonstrated range resolution is its distinction from the depth resolution in ultrasound imaging, which is known to be dependent on the detection frequency and the corresponding wavelength. The depth (or axial) resolution in ultrasound imaging at 500 kHz is about 3 mm, which is much worse than the submillimeter resolution typically featured in these ionoacoustic experiments. In contrast to ultrasound imaging, which attempts to resolve two objects with a certain spatial separation (limited by the wavelength), the location of the BP maximum is determined in ionoacoustic measurements, where resolution is defined by the time resolution of the detection system and exhibits a weak frequency dependence only [17].

Particle detectors such as semiconductor devices, ionization chambers or scintillators often suffer from saturation effects at very high ion bunch intensities. Acoustic detection of particles is based on the ion energy deposition in the detector medium and its conversion to heat. Thus, as long as no phase change is induced in water and within stress confinement, the pressure signal delivers the ion range or energy with an amplitude that is linearly proportional to the number of ions in the bunch. The total energy stored in the BPV (i.e. deposited dose) defines the temperature and corresponding pressure increase. These macroscopic physical parameters are determined on the one hand by the ion intensity, energy and nuclear charge, and on the other hand by the lateral and axial dimensions of the BPV. These values can be calculated in Geant4 simulations to determine the temperature and pressure within the BPV. One can estimate the ionoacoustic pressure by $\Delta p = \beta \Delta T / \kappa$ for ions used in this work. Assuming $10^6$ ions per pulse within a beam spot of 1 mm$^2$ and the corresponding BP width, this yields 7 Pa for 20 MeV protons, 12 Pa for 200 MeV/u $^{12}$C, 3.5 kPa for 300 MeV/u $^{124}$Xe and 4.8 kPa for 300 MeV/u $^{138}$U ions. The measured signal amplitude at the detector position depends on the detection geometry, the amplification and the detector transmission function TIR. The lowest pulse intensity, which was measured with our standard setup was $10^4$ for protons and $10^2$ for U ions using 60 dB amplification. When lower amplification and signal attenuation (via larger BP-transducer distance) are also considered, a dynamic range that spans several orders of magnitude seems quite feasible.

To make use of the ionoacoustic method for pulsed heavy ion beam monitoring a next step would be to develop a more compact detection design with a water volume adjusted to the expected ion range and a transducer stationary that is mounted in the back wall of the detector housing. A tailored compact monitor of this sort can be calibrated to a certain ion species for energy measurements. It has been demonstrated that even the energy distribution of a single ion bunch can be reconstructed from the ionoacoustic signal shape using a novel technique (called I-BEAT), that makes further use of the detector transfer function [12]. To replace an accelerator trigger, an appropriate scintillation detector can be used in transmission or attached to the monitor looking for prompt reaction gammas [9]. Although convenient, water is not the most sensitive detection medium. To enhance the sensitivity different liquids with larger $\beta/\kappa$ ratios have been considered and tested in the past [1, 33]. Extension of this 1D configuration to 3D monitoring would afford the simultaneous measurement of beam position, ion energy (i.e. bunch spectrum) and 3D dose distribution. First tests with 20 MeV protons using one axial and three lateral transducers showed submillimeter accuracy for the beam position and its lateral extension.

Stopping power measurements are an obvious application example of using ionoacoustics to achieve unrivaled simple, fast and efficient detection with high accuracy. All other energy loss methods in this high energy range can require bulky spectrometers and data acquisition systems, which can deliver more detailed information, but often only energy loss and energy loss straggling are of interest [34]. Here, after a range-energy calibration (an example is shown for $^{124}$Xe in Fig. 12), materials of interest with appropriate thicknesses could be mounted in front of the detector, e.g. on a target wheel and changed by remote control after each measurement. The specific energy loss and straggling values are obtained immediately from the acquired signal. With a 3D detector configuration, even lateral scattering can be reconstructed from transducer data. In addition to particle range in water, the water equivalent thickness (WET) of different materials is of significant interest in hadron therapy. This can easily and precisely be measured with this technique which is noteworthy for heavy ions [17].

In summary, notwithstanding the simplicity of the acoustic particle detector, our experimental tests have demonstrated the great potential of the ionoacoustic method for monitoring heavy ion beams. Unlike electronic particle detectors, acoustic transducers are insensitive to gamma and neutron radiation (and the associated deleterious background they can generate) and the detector medium itself is not affected by radiation damage. Given the high precision and accuracy of this method as well as its capacity for a huge dynamic range, ionoacoustic detection has the prospect to become a standard tool at accelerator facilities offering short bunches of swift heavy ions up to highest intensities.

## Acknowledgments

The authors would like to thank Marco Pinto, Franz S. Englbrecht, Julie Lascaud , Andreas Maaß and, in particular, Paul Bolton for valuable contributions. Dieter Schardt (GSI) is kindly acknowledged for sharing unpublished data. The results presented here are based on experiments, which were performed at the beam line HTB at the GSI Helmholtzzentrum für Schwerionenforschung, Darmstadt (Germany) in the frame of FAIR Phase-0.We gratefully acknowledge the support and excellent beam quality delivered by accelerator staff members at both GSI and MLL. This work was funded by the DFG Cluster of Excellence Munich Centre for Advanced Photonics (MAP).